\documentclass[aps, pre, twocolumn, amsmath, amssymb, floatfix, showpacs]{revtex4}
\usepackage[english]{babel}
\usepackage[latin2]{inputenc}
\usepackage[T1]{fontenc}

\usepackage[usenames]{color}
\usepackage{ae,aecompl}
\usepackage{enumerate}
\usepackage{epsfig}
\usepackage{graphics}
\usepackage{graphicx}
\usepackage{amsmath}
\usepackage{amssymb}
\usepackage{pifont}
\usepackage[sort&compress]{natbib}
\usepackage{wasysym}
\usepackage{color}
\usepackage{times}

\def\ev #1{\left\langle #1 \right\rangle}
\def\br #1{\left( #1 \right)}
\def\Br #1{\left[ #1 \right]}
\def\BR #1{\left\{ #1 \right\}}
\def\abs #1{\left| #1 \right|}
\def\prob #1{\mathbb P \left( #1 \right)}

\newbox\bwk\edef\tempd#1pt{#1\string p\string t}\tempd\def\nbextr#1pt{#1}
\def\npts#1{\expandafter\nbextr\the#1\space}
\def\ttwplink#1#2{
#2
\setbox\bwk=\hbox{#2}\special{ps:( linkto #1)\space\npts{\wd\bwk} \npts{\dp\bwk} -\npts{\ht\bwk} true\space Cpos}}

\begin{document}

\title{A shift-optimized Hill-type estimator}
\author{\'Eva R\'acz}
\affiliation{Department of Theoretical Physics, Budapest University of Technology and Economics, Budapest, Hungary}
\author{J\'anos Kert\'esz}
\affiliation{Department of Theoretical Physics, Budapest University of Technology and Economics, Budapest, Hungary}
\author{Zolt\'an Eisler}
\affiliation{Nimbus Volatility Arbitrage, Capital Fund Management, Paris, France}
\date{\today}

\begin{abstract}
A wide range of natural and social phenomena result in observables whose distributions can be well approximated by a power-law decay.
 The well-known Hill estimator of the tail exponent provides results which are in many respects superior to other estimators in case the asymptotics of the distribution is indeed a pure power-law, however, systematic errors occur if the distribution is altered by simply shifting it.
 We demonstrate some related problems which typically emerge when dealing with empirical data and suggest a procedure designed to extend the applicability of the Hill estimator.
\end{abstract}

\pacs{02.50.Tt, 05.45.Tp, 89.65.Gh}
\maketitle

\section{Introduction}
	\label{sec:Introduction}
%1. bekezdes: Mirol is beszelunk?
Heavy-tailed distributions emerge in many situations, with examples ranging from social networks to earthquake intensities, city sizes etc. (for further examples, see \cite{Newman05} and references therein). In mathematical terms, a random variable $X$ has a heavy upper tail if the probability
\begin{equation}
 \mathbb P \br{X \geq x} \propto x^{-\alpha} L \br x,
 \label{eq:heavytails}
\end{equation}
with $\alpha >0$ and $L\br x$ being a slowly varying function of its argument.
 The function $\bar F \br x \equiv \prob{X \geqslant x}$ is called the (complementary) cumulative distribution function (cdf in the following).
 The exponent $\alpha$, termed the \emph{tail exponent} is a parameter of practical importance, since this is the quantitity which determines the frequency of extreme events (e.\ g.\ huge losses on the stock market).

%2. bekezdes: A feladat kituzese
The general problem can be formulated as follows: given some finite sample $\mathcal S= \BR{X_1, X_2,\dots, X_N}$ of independent, identically distributed elements, of which the distribution can be described by Eq.\ \eqref{eq:heavytails}, we intend to find an efficient procedure to estimate the tail exponent.
 There exist many such estimators, each of those has its advantages and drawbacks. The difficulties lie generally in the following:
\begin{itemize}
	\item The small $x$ form of the cdf, which in Eq.\ \eqref{eq:heavytails} is incorporated in $L\br x$, can shorten the ``effective tail length'', i.\ e., the domain where the distribution is close to a power-law. Therefore, the actual form of $L\br{x}$ affects the speed of convergence of any estimator.
	\item If $\alpha$ is relatively large, one needs a huge dataset to have enough points in the tail.
	\item Linear transformations of a random variable do not affect its asymptotic behavior (i.\ e., $\alpha$), but can affect the value of an estimator over a finite sample.
\end{itemize}

%3. bekezdes: A Hill becsles
The popular Hill estimator \cite{Hill75} (HE in the following) is based on the $n$ largest observations in the sample, and is defined as follows:
	\begin{equation}
	  \hat\alpha_{\mathrm H}\br{\mathcal S, n} \equiv \Br{\frac{1}{n-1}\mathop{\sum}_{j=1}^{n-1}\ln\br{\frac{X_{\br j}}{X_{\br n}}}}^{-1},
	  \label{eq:HillEstimator}
	\end{equation}
with $X_{\br 1} \geq X_{\br 2} \geq \dots \geq X_{\br N}$ being elements of the order statistics.
 Note that Eq.\ \eqref{eq:HillEstimator} is invariant to multiplication ($\hat\alpha_{\mathrm H}\br{a\cdot \mathcal S, n} = \hat\alpha_{\mathrm H}\br{\mathcal S, n}$), yet not shift-invariant ($\hat\alpha_{\mathrm H}\br{\mathcal S +s, n} \neq \hat \alpha_{\mathrm H}\br{\mathcal S, n}$).
 For a fixed $n$, the HE is a maximum likelihood estimator of the tail exponent, but the appropriate choice of the tail length $n$ remains an issue, since $\hat \alpha_{\mathrm H}$ is typically very sensitive to it. The standard way to determine the threshold $x_0 \equiv X_{\br n}$ is to construct a so-called Hill plot, which is $1/\hat\alpha_{\mathrm H}$ as a function of $n$, and look for a plateau in the graph (for other evaluation methods, see \cite{Drees00}).

%4. bekezdes: CSNE
In a recent publication \cite{Clauset07}, Clauset \emph{et al.} suggest a procedure (CSNE in the following) to solve the former problem, i.\ e.\ to find the optimal $n$, in an automated fashion. This method provides superior results if $L\br x = \mathrm{const.}$, but inherits the sensitivity of the Hill estimator to the actual form of $L\br x$. 

The shifted Pareto tail (Eq.\ \eqref{eq:ShiftedPowerLaw}) presents a special case of Eq.\ \eqref{eq:heavytails}:
	\begin{equation}
		\bar F \br x \propto \br{x+s}^{-\alpha} \equiv x^{-\alpha}\cdot \br{1+s/x}^{-\alpha}.
		\label{eq:ShiftedPowerLaw}
	\end{equation}
Although, at first sight, the introduction of data shifts does not seem a drastic change, it limits the applicability of both the HE and the CSNE procedures.

A method to tackle the problem of data shifts is for example the Fraga Alves estimator \cite{FragaAlves01}, which is invariant to both shifts and multiplication of the dataset, at the cost of a slow convergence. Another example is the Meerschaert--Scheffler estimator \cite{Meerschaert98} which is shift-independent, but not invariant to multiplication, and furthermore its applicability is restricted to $\alpha < 2$.

The aim of the present work is to show an extension of the Hill estimator which can handle both the threshold and the shift problem, in a similar procedure as CSNE. The paper is organized as follows: 
In Section \ref{sec:OptimizingTheShift} we analyze the systematic errors introduced by the shift in the distribution and demonstrate how this can be taken into account in the estimator. Section \ref{sec:SimulationResults} demonstrates the performance of the suggested method on computer-generated data. In Section \ref{sec:EmpiricalData} we present an analysis of empirical data as taken from traded volumes of stocks on the stock market. In Section \ref{sec:Conclusions} we give the conclusions, the Appendix briefly summarizes the CSNE algorithm \cite{Clauset07}.

\begin{figure}[tb]
        \centering
       %\input{ntailCF_1.tex}
% GNUPLOT: LaTeX picture with Postscript
\begingroup
  \makeatletter
  \providecommand\color[2][]{%
    \GenericError{(gnuplot) \space\space\space\@spaces}{%
      Package color not loaded in conjunction with
      terminal option `colourtext'%
    }{See the gnuplot documentation for explanation.%
    }{Either use 'blacktext' in gnuplot or load the package
      color.sty in LaTeX.}%
    \renewcommand\color[2][]{}%
  }%
  \providecommand\includegraphics[2][]{%
    \GenericError{(gnuplot) \space\space\space\@spaces}{%
      Package graphicx or graphics not loaded%
    }{See the gnuplot documentation for explanation.%
    }{The gnuplot epslatex terminal needs graphicx.sty or graphics.sty.}%
    \renewcommand\includegraphics[2][]{}%
  }%
  \providecommand\rotatebox[2]{#2}%
  \@ifundefined{ifGPcolor}{%
    \newif\ifGPcolor
    \GPcolortrue
  }{}%
  \@ifundefined{ifGPblacktext}{%
    \newif\ifGPblacktext
    \GPblacktexttrue
  }{}%
  % define a \g@addto@macro without @ in the name:
  \let\gplgaddtomacro\g@addto@macro
  % define empty templates for all commands taking text:
  \gdef\gplbacktext{}%
  \gdef\gplfronttext{}%
  \makeatother
  \ifGPblacktext
    % no textcolor at all
    \def\colorrgb#1{}%
    \def\colorgray#1{}%
  \else
    % gray or color?
    \ifGPcolor
      \def\colorrgb#1{\color[rgb]{#1}}%
      \def\colorgray#1{\color[gray]{#1}}%
      \expandafter\def\csname LTw\endcsname{\color{white}}%
      \expandafter\def\csname LTb\endcsname{\color{black}}%
      \expandafter\def\csname LTa\endcsname{\color{black}}%
      \expandafter\def\csname LT0\endcsname{\color[rgb]{1,0,0}}%
      \expandafter\def\csname LT1\endcsname{\color[rgb]{0,1,0}}%
      \expandafter\def\csname LT2\endcsname{\color[rgb]{0,0,1}}%
      \expandafter\def\csname LT3\endcsname{\color[rgb]{1,0,1}}%
      \expandafter\def\csname LT4\endcsname{\color[rgb]{0,1,1}}%
      \expandafter\def\csname LT5\endcsname{\color[rgb]{1,1,0}}%
      \expandafter\def\csname LT6\endcsname{\color[rgb]{0,0,0}}%
      \expandafter\def\csname LT7\endcsname{\color[rgb]{1,0.3,0}}%
      \expandafter\def\csname LT8\endcsname{\color[rgb]{0.5,0.5,0.5}}%
    \else
      % gray
      \def\colorrgb#1{\color{black}}%
      \def\colorgray#1{\color[gray]{#1}}%
      \expandafter\def\csname LTw\endcsname{\color{white}}%
      \expandafter\def\csname LTb\endcsname{\color{black}}%
      \expandafter\def\csname LTa\endcsname{\color{black}}%
      \expandafter\def\csname LT0\endcsname{\color{black}}%
      \expandafter\def\csname LT1\endcsname{\color{black}}%
      \expandafter\def\csname LT2\endcsname{\color{black}}%
      \expandafter\def\csname LT3\endcsname{\color{black}}%
      \expandafter\def\csname LT4\endcsname{\color{black}}%
      \expandafter\def\csname LT5\endcsname{\color{black}}%
      \expandafter\def\csname LT6\endcsname{\color{black}}%
      \expandafter\def\csname LT7\endcsname{\color{black}}%
      \expandafter\def\csname LT8\endcsname{\color{black}}%
    \fi
  \fi
  \setlength{\unitlength}{0.0500bp}%
  \begin{picture}(4320.00,3022.00)%
    \gplgaddtomacro\gplbacktext{%
      \csname LTb\endcsname%
      \put(780,600){\makebox(0,0)[r]{\strut{}1e1}}%
      \put(780,1327){\makebox(0,0)[r]{\strut{}1e2}}%
      \put(780,2055){\makebox(0,0)[r]{\strut{}1e3}}%
      \put(780,2782){\makebox(0,0)[r]{\strut{}1e4}}%
      \put(900,400){\makebox(0,0){\strut{}-3}}%
      \put(1413,400){\makebox(0,0){\strut{}-2}}%
      \put(1927,400){\makebox(0,0){\strut{}-1}}%
      \put(2440,400){\makebox(0,0){\strut{} 0}}%
      \put(2953,400){\makebox(0,0){\strut{} 1}}%
      \put(3467,400){\makebox(0,0){\strut{} 2}}%
      \put(3980,400){\makebox(0,0){\strut{} 3}}%
      \put(200,1691){\rotatebox{90}{\makebox(0,0){\strut{}Estimated tail length (CSNE)}}}%
      \put(2440,100){\makebox(0,0){\strut{}Shift (rel. units)}}%
    }%
    \gplgaddtomacro\gplfronttext{%
    }%
    \gplbacktext
    \put(0,0){\includegraphics{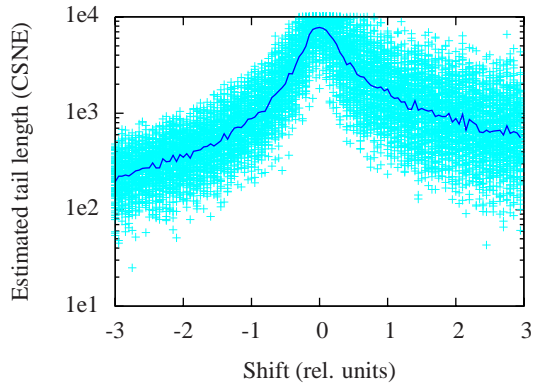}}%
    \gplfronttext
  \end{picture}%
\endgroup

        \caption{Effective tail length as a function of data shift. Data details: $\alpha = 2.36$ pure power-law, shifts are given in units of the mean absolute deviation ($\mathbb E \br {\left| X - \mathbb E \br X \right|} = \frac{2}{\alpha -1}(\frac{\alpha}{\alpha - 1})^{-\alpha+1} \approx 0.7$). For each shift, 100 samples of size 10000 were generated, the cyan colored points correspond to the result of the fitting procedure on each of those. The blue colored curve corresponds to the mean of the 100 runs at each shift. (color online)}
        \label{fig:scn236}
\end{figure}

\begin{figure}[tb]
        \centering
                %\input{expCF.tex}
% GNUPLOT: LaTeX picture with Postscript
\begingroup
  \makeatletter
  \providecommand\color[2][]{%
    \GenericError{(gnuplot) \space\space\space\@spaces}{%
      Package color not loaded in conjunction with
      terminal option `colourtext'%
    }{See the gnuplot documentation for explanation.%
    }{Either use 'blacktext' in gnuplot or load the package
      color.sty in LaTeX.}%
    \renewcommand\color[2][]{}%
  }%
  \providecommand\includegraphics[2][]{%
    \GenericError{(gnuplot) \space\space\space\@spaces}{%
      Package graphicx or graphics not loaded%
    }{See the gnuplot documentation for explanation.%
    }{The gnuplot epslatex terminal needs graphicx.sty or graphics.sty.}%
    \renewcommand\includegraphics[2][]{}%
  }%
  \providecommand\rotatebox[2]{#2}%
  \@ifundefined{ifGPcolor}{%
    \newif\ifGPcolor
    \GPcolortrue
  }{}%
  \@ifundefined{ifGPblacktext}{%
    \newif\ifGPblacktext
    \GPblacktexttrue
  }{}%
  % define a \g@addto@macro without @ in the name:
  \let\gplgaddtomacro\g@addto@macro
  % define empty templates for all commands taking text:
  \gdef\gplbacktext{}%
  \gdef\gplfronttext{}%
  \makeatother
  \ifGPblacktext
    % no textcolor at all
    \def\colorrgb#1{}%
    \def\colorgray#1{}%
  \else
    % gray or color?
    \ifGPcolor
      \def\colorrgb#1{\color[rgb]{#1}}%
      \def\colorgray#1{\color[gray]{#1}}%
      \expandafter\def\csname LTw\endcsname{\color{white}}%
      \expandafter\def\csname LTb\endcsname{\color{black}}%
      \expandafter\def\csname LTa\endcsname{\color{black}}%
      \expandafter\def\csname LT0\endcsname{\color[rgb]{1,0,0}}%
      \expandafter\def\csname LT1\endcsname{\color[rgb]{0,1,0}}%
      \expandafter\def\csname LT2\endcsname{\color[rgb]{0,0,1}}%
      \expandafter\def\csname LT3\endcsname{\color[rgb]{1,0,1}}%
      \expandafter\def\csname LT4\endcsname{\color[rgb]{0,1,1}}%
      \expandafter\def\csname LT5\endcsname{\color[rgb]{1,1,0}}%
      \expandafter\def\csname LT6\endcsname{\color[rgb]{0,0,0}}%
      \expandafter\def\csname LT7\endcsname{\color[rgb]{1,0.3,0}}%
      \expandafter\def\csname LT8\endcsname{\color[rgb]{0.5,0.5,0.5}}%
    \else
      % gray
      \def\colorrgb#1{\color{black}}%
      \def\colorgray#1{\color[gray]{#1}}%
      \expandafter\def\csname LTw\endcsname{\color{white}}%
      \expandafter\def\csname LTb\endcsname{\color{black}}%
      \expandafter\def\csname LTa\endcsname{\color{black}}%
      \expandafter\def\csname LT0\endcsname{\color{black}}%
      \expandafter\def\csname LT1\endcsname{\color{black}}%
      \expandafter\def\csname LT2\endcsname{\color{black}}%
      \expandafter\def\csname LT3\endcsname{\color{black}}%
      \expandafter\def\csname LT4\endcsname{\color{black}}%
      \expandafter\def\csname LT5\endcsname{\color{black}}%
      \expandafter\def\csname LT6\endcsname{\color{black}}%
      \expandafter\def\csname LT7\endcsname{\color{black}}%
      \expandafter\def\csname LT8\endcsname{\color{black}}%
    \fi
  \fi
  \setlength{\unitlength}{0.0500bp}%
  \begin{picture}(4320.00,3022.00)%
    \gplgaddtomacro\gplbacktext{%
      \csname LTb\endcsname%
      \put(780,600){\makebox(0,0)[r]{\strut{}1}}%
      \put(780,873){\makebox(0,0)[r]{\strut{}1.5}}%
      \put(780,1146){\makebox(0,0)[r]{\strut{}2}}%
      \put(780,1418){\makebox(0,0)[r]{\strut{}2.5}}%
      \put(780,1691){\makebox(0,0)[r]{\strut{}3}}%
      \put(780,1964){\makebox(0,0)[r]{\strut{}3.5}}%
      \put(780,2237){\makebox(0,0)[r]{\strut{}4}}%
      \put(780,2509){\makebox(0,0)[r]{\strut{}4.5}}%
      \put(780,2782){\makebox(0,0)[r]{\strut{}5}}%
      \put(900,400){\makebox(0,0){\strut{}-3}}%
      \put(1413,400){\makebox(0,0){\strut{}-2}}%
      \put(1927,400){\makebox(0,0){\strut{}-1}}%
      \put(2440,400){\makebox(0,0){\strut{} 0}}%
      \put(2953,400){\makebox(0,0){\strut{} 1}}%
      \put(3467,400){\makebox(0,0){\strut{} 2}}%
      \put(3980,400){\makebox(0,0){\strut{} 3}}%
      \put(200,1691){\rotatebox{90}{\makebox(0,0){\strut{}Estimated exponent (CSNE)}}}%
      \put(2440,100){\makebox(0,0){\strut{}Shift (rel. units)}}%
    }%
    \gplgaddtomacro\gplfronttext{%
      \csname LTb\endcsname%
      \put(1764,2573){\makebox(0,0)[l]{\strut{}Correct value (2.36)}}%
      \csname LTb\endcsname%
      \put(1764,2373){\makebox(0,0)[l]{\strut{}Mean}}%
    }%
    \gplbacktext
    \put(0,0){\includegraphics{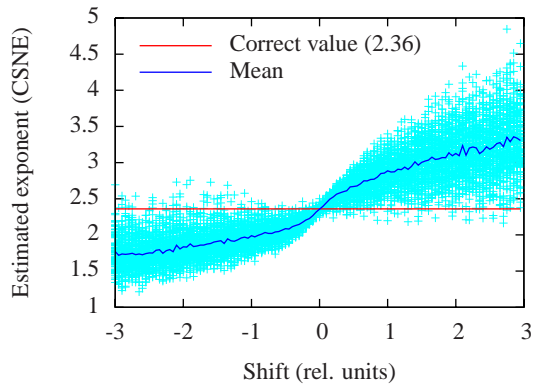}}%
    \gplfronttext
  \end{picture}%
\endgroup

        \caption{Estimated exponents as a function of data shift, same setup as in Fig.\ \ref{fig:scn236}. The blue colored curve corresponds again to the average of the estimates. Note that shifts result not only in a shorter tail, thus a larger standard deviation, but also in the shift of the mean estimate. (color online)}
        \label{fig:sca236}
\end{figure}

\section{Optimizing the shift}
	\label{sec:OptimizingTheShift}
In short, the CSNE algorithm \cite{Clauset07} optimizes the parameter $n$ of Hill's estimator, so that the distance of the fitted power-law and the conditional cdf be minimal (for a summary in terms of formulas, see the Appendix). Clauset~\emph{et al.}\ suggest the Kolmogorov--Smirnov (KS in the following) statistic ($D_{\mathrm{KS}}\br{F,G} \equiv \sup _x\abs{F\br x - G\br x}$) as the definition of distance. This method has proven to be a useful tool, however, tests on shifted power-law samples (Eq.\ \eqref{eq:ShiftedPowerLaw}) show that in that case it provides biased results (Figs.\ \ref{fig:scn236} and \ref{fig:sca236}).
 While shifts are irrelevant asymptotically (if $\abs{s} \ll x$, then $\br{x+s}^{-\alpha} \approx x^{-\alpha}$), having a finite dataset at hand they result in a shorter ``effective tail length'' (i.\ e.\ the threshold above which the $\br{x+s}^{-\alpha} \approx x^{-\alpha}$ approximation is valid becomes higher). This observation explains that with growing shift, the CSNE procedure provides more and more volatile results (Fig.\ \ref{fig:sca236}). The average estimate deviates from the zero shift value since in the shifted case, the cdf is deterministically either convex or concave, i.\ e.\, the estimator is bound to deviate from the true value in a fixed direction.
 Thus, the task is to optimize the tail length $n$ and the shift parameter $s$ simultaneously. If this is achieved, the Hill formula can be applied to the $n$ largest observations shifted with the previously obtained value of $s$.

In the simplest case, let us assume that whole the sample is taken from a shifted power-law distribution, i.\ e., we do not have to deal with estimating the tail length. Our aim is to find a shift estimator $\hat s\br{\mathcal S}$, for which $\mathcal S' = \mathcal S + \hat s\br{\mathcal S}$ is well-approximated by a pure, non-shifted Pareto law. Note that, from the practitioner's point of view, $\hat s$ does not necessarily need to be very accurate, since as Fig.\ \ref{fig:sca236} shows, the mean estimate depends smoothly on the shift and has a small standard deviation in the vicinity of zero shift.

\begin{figure}[htb]
	\centering
		%\input{shiftest-2.0.tex}
% GNUPLOT: LaTeX picture with Postscript
\begingroup
  \makeatletter
  \providecommand\color[2][]{%
    \GenericError{(gnuplot) \space\space\space\@spaces}{%
      Package color not loaded in conjunction with
      terminal option `colourtext'%
    }{See the gnuplot documentation for explanation.%
    }{Either use 'blacktext' in gnuplot or load the package
      color.sty in LaTeX.}%
    \renewcommand\color[2][]{}%
  }%
  \providecommand\includegraphics[2][]{%
    \GenericError{(gnuplot) \space\space\space\@spaces}{%
      Package graphicx or graphics not loaded%
    }{See the gnuplot documentation for explanation.%
    }{The gnuplot epslatex terminal needs graphicx.sty or graphics.sty.}%
    \renewcommand\includegraphics[2][]{}%
  }%
  \providecommand\rotatebox[2]{#2}%
  \@ifundefined{ifGPcolor}{%
    \newif\ifGPcolor
    \GPcolortrue
  }{}%
  \@ifundefined{ifGPblacktext}{%
    \newif\ifGPblacktext
    \GPblacktexttrue
  }{}%
  % define a \g@addto@macro without @ in the name:
  \let\gplgaddtomacro\g@addto@macro
  % define empty templates for all commands taking text:
  \gdef\gplbacktext{}%
  \gdef\gplfronttext{}%
  \makeatother
  \ifGPblacktext
    % no textcolor at all
    \def\colorrgb#1{}%
    \def\colorgray#1{}%
  \else
    % gray or color?
    \ifGPcolor
      \def\colorrgb#1{\color[rgb]{#1}}%
      \def\colorgray#1{\color[gray]{#1}}%
      \expandafter\def\csname LTw\endcsname{\color{white}}%
      \expandafter\def\csname LTb\endcsname{\color{black}}%
      \expandafter\def\csname LTa\endcsname{\color{black}}%
      \expandafter\def\csname LT0\endcsname{\color[rgb]{1,0,0}}%
      \expandafter\def\csname LT1\endcsname{\color[rgb]{0,1,0}}%
      \expandafter\def\csname LT2\endcsname{\color[rgb]{0,0,1}}%
      \expandafter\def\csname LT3\endcsname{\color[rgb]{1,0,1}}%
      \expandafter\def\csname LT4\endcsname{\color[rgb]{0,1,1}}%
      \expandafter\def\csname LT5\endcsname{\color[rgb]{1,1,0}}%
      \expandafter\def\csname LT6\endcsname{\color[rgb]{0,0,0}}%
      \expandafter\def\csname LT7\endcsname{\color[rgb]{1,0.3,0}}%
      \expandafter\def\csname LT8\endcsname{\color[rgb]{0.5,0.5,0.5}}%
    \else
      % gray
      \def\colorrgb#1{\color{black}}%
      \def\colorgray#1{\color[gray]{#1}}%
      \expandafter\def\csname LTw\endcsname{\color{white}}%
      \expandafter\def\csname LTb\endcsname{\color{black}}%
      \expandafter\def\csname LTa\endcsname{\color{black}}%
      \expandafter\def\csname LT0\endcsname{\color{black}}%
      \expandafter\def\csname LT1\endcsname{\color{black}}%
      \expandafter\def\csname LT2\endcsname{\color{black}}%
      \expandafter\def\csname LT3\endcsname{\color{black}}%
      \expandafter\def\csname LT4\endcsname{\color{black}}%
      \expandafter\def\csname LT5\endcsname{\color{black}}%
      \expandafter\def\csname LT6\endcsname{\color{black}}%
      \expandafter\def\csname LT7\endcsname{\color{black}}%
      \expandafter\def\csname LT8\endcsname{\color{black}}%
    \fi
  \fi
  \setlength{\unitlength}{0.0500bp}%
  \begin{picture}(5040.00,3528.00)%
    \gplgaddtomacro\gplbacktext{%
      \csname LTb\endcsname%
      \put(990,860){\makebox(0,0)[r]{\strut{}-0.2}}%
      \put(990,1261){\makebox(0,0)[r]{\strut{}-0.1}}%
      \put(990,1662){\makebox(0,0)[r]{\strut{} 0}}%
      \put(990,2062){\makebox(0,0)[r]{\strut{} 0.1}}%
      \put(990,2463){\makebox(0,0)[r]{\strut{} 0.2}}%
      \put(990,2863){\makebox(0,0)[r]{\strut{} 0.3}}%
      \put(990,3264){\makebox(0,0)[r]{\strut{} 0.4}}%
      \put(1122,440){\makebox(0,0){\strut{}-1}}%
      \put(1476,440){\makebox(0,0){\strut{}-0.8}}%
      \put(1831,440){\makebox(0,0){\strut{}-0.6}}%
      \put(2185,440){\makebox(0,0){\strut{}-0.4}}%
      \put(2540,440){\makebox(0,0){\strut{}-0.2}}%
      \put(2894,440){\makebox(0,0){\strut{} 0}}%
      \put(3248,440){\makebox(0,0){\strut{} 0.2}}%
      \put(3603,440){\makebox(0,0){\strut{} 0.4}}%
      \put(3957,440){\makebox(0,0){\strut{} 0.6}}%
      \put(4312,440){\makebox(0,0){\strut{} 0.8}}%
      \put(4666,440){\makebox(0,0){\strut{} 1}}%
      \put(220,1962){\rotatebox{90}{\makebox(0,0){\strut{}Error of estimated shift}}}%
      \put(2894,110){\makebox(0,0){\strut{}True shift}}%
    }%
    \gplgaddtomacro\gplfronttext{%
      \csname LTb\endcsname%
      \put(3679,3091){\makebox(0,0)[r]{\strut{}$\Delta \hat s$}}%
      \csname LTb\endcsname%
      \put(3679,2871){\makebox(0,0)[r]{\strut{}$\ev{\Delta \hat s}$}}%
      \csname LTb\endcsname%
      \put(3679,2651){\makebox(0,0)[r]{\strut{}$\ev{\Delta \hat s}\pm\sqrt{\ev{\Delta \hat s^2}}$}}%
    }%
    \gplbacktext
    \put(0,0){\includegraphics{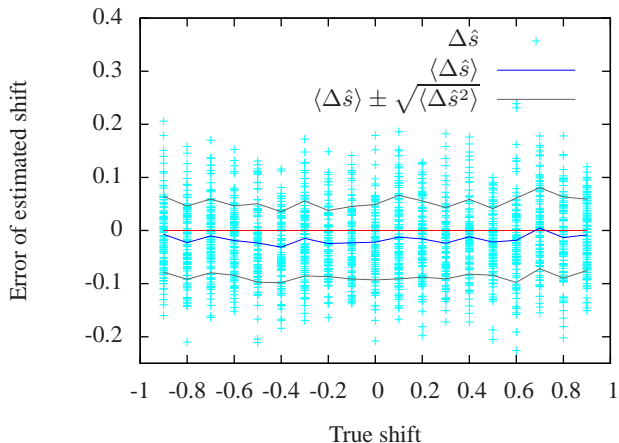}}%
    \gplfronttext
  \end{picture}%
\endgroup

		\caption{The figure shows the shift estimator's output on computer-generated shifted power-law samples ($\alpha = 2.0$, $N=10^4$), at each shift value, 100 trials were performed. Note that the procedure is definitely shift-invariant, as intended. (color online)}
	\label{fig:linregestimator}
\end{figure}

In geometric terms, we have to ``straighten out'' the cdf plot, i.\ e., to determine the shift so that the cdf of $\mathcal S'$ on a doubly logarithmic plot is as close to a straight line as possible.
 The simplest way to achieve this is to minimize the mean squared error of the linear fit on the log-log plot (via numerical optimization, e.\ g.\ the golden section method \cite{NumericalRecipes}).
 Figure \ref{fig:linregestimator} and Table \ref{tab:linregestimator} show the performance of the latter procedure on computer-generated shifted Pareto samples. It can be concluded that although this type of estimator slightly underestimates the shift on average, it provides reasonable results.

\begin{table}[htb]
	\centering
	\begin{tabular}{c|c|c}
		$\alpha$ & $\ev{\Delta \hat s}$ & $\sqrt{\ev{\Delta \hat s^2}}$ \\
		\hline
		\hline
		1.5 & -0.018 & 0.070 \\
		1.75 & -0.019 & 0.070 \\
		2.0 & -0.020 & 0.072 \\
		2.25 & -0.021 & 0.074 \\
		2.5 & -0.023 & 0.076 \\
	\end{tabular}
	\caption{As the table shows, the procedure depends on $\alpha$, its bias and standard deviation both increase slightly with growing $\alpha$. (The averages were taken over all shifts for a fixed exponent, i. e. 19x1000 trials with sample size $N = 10^4$.)}
	\label{tab:linregestimator}
\end{table}

Back to the general case, where the tail length can be smaller than the sample size, one has to estimate $n$ along with the shift $s$.
A consistent and simple way to incorporate the shift estimator in a procedure in the manner of \cite{Clauset07} is to estimate at each tail length $n$ the shift $s$ based on the $n$ largest entries of the sample. Having obtained $s$, one has to shift the tail with this value, and calculate the Hill estimator using the shifted tail. Thus, we obtain $\br{n_i, s_i, \alpha_i}$, $i = 1,2,\ldots,K\leqslant N$, and from these, we accept the one which is the closest to the empirical cdf, according to the KS statistic. So the ``recipe'' is the following:
\begin{enumerate}
 \item For each tail length $n$, calculate $\hat s \br{\mathcal T_n}$ (with $\mathcal T_n$ denoting the $n$ largest elements in the sample),
 \item calculate $\hat \alpha_{\mathrm H}\br{\mathcal S + \hat s \br{\mathcal T_n}, n}$,
 \item calculate the KS-distance between the cdf of $\mathcal T_n$ and $\br{\frac{x+\hat s}{x_0 + \hat s}}^{-\alpha_{\mathrm H}}$, with $x_0 \equiv X_{\br n}$, as previously.
 \item Accept the fit with the lowest KS-statistic.
\end{enumerate}
The set of $n$ tail lengths to test is chosen in the same manner as in the CSNE procedure (see the Appendix).
Sections \ref{sec:SimulationResults} and \ref{sec:EmpiricalData} analyze the capabilities of this procedure on computer-generated and empirical data.
When considering empirical data, one cannot assume that the cdf has exactly the form of Eq.\ \eqref{eq:ShiftedPowerLaw}, rather a variant of Eq.\ \eqref{eq:heavytails}:
\begin{equation}
 \bar F \br x \propto \br{x + s}^{-\alpha} \cdot \tilde L\br x \equiv x^{-\alpha} \cdot L\br x.
 \label{eq:heavytails2}
\end{equation}
Although the difference between Eq.\ \eqref{eq:heavytails} and \eqref{eq:heavytails2} is only in grouping, it can pay off in case $\tilde L\br x$ is closer to a constant than $L \br x$.

\section{Simulation results}
	\label{sec:SimulationResults}
\begin{figure}[tb]
 	\centering
		%\input{1.50-test_shfit-1.exp.tex}
% GNUPLOT: LaTeX picture with Postscript
\begingroup
  \makeatletter
  \providecommand\color[2][]{%
    \GenericError{(gnuplot) \space\space\space\@spaces}{%
      Package color not loaded in conjunction with
      terminal option `colourtext'%
    }{See the gnuplot documentation for explanation.%
    }{Either use 'blacktext' in gnuplot or load the package
      color.sty in LaTeX.}%
    \renewcommand\color[2][]{}%
  }%
  \providecommand\includegraphics[2][]{%
    \GenericError{(gnuplot) \space\space\space\@spaces}{%
      Package graphicx or graphics not loaded%
    }{See the gnuplot documentation for explanation.%
    }{The gnuplot epslatex terminal needs graphicx.sty or graphics.sty.}%
    \renewcommand\includegraphics[2][]{}%
  }%
  \providecommand\rotatebox[2]{#2}%
  \@ifundefined{ifGPcolor}{%
    \newif\ifGPcolor
    \GPcolortrue
  }{}%
  \@ifundefined{ifGPblacktext}{%
    \newif\ifGPblacktext
    \GPblacktextfalse
  }{}%
  % define a \g@addto@macro without @ in the name:
  \let\gplgaddtomacro\g@addto@macro
  % define empty templates for all commands taking text:
  \gdef\gplbacktext{}%
  \gdef\gplfronttext{}%
  \makeatother
  \ifGPblacktext
    % no textcolor at all
    \def\colorrgb#1{}%
    \def\colorgray#1{}%
  \else
    % gray or color?
    \ifGPcolor
      \def\colorrgb#1{\color[rgb]{#1}}%
      \def\colorgray#1{\color[gray]{#1}}%
      \expandafter\def\csname LTw\endcsname{\color{white}}%
      \expandafter\def\csname LTb\endcsname{\color{black}}%
      \expandafter\def\csname LTa\endcsname{\color{black}}%
      \expandafter\def\csname LT0\endcsname{\color[rgb]{1,0,0}}%
      \expandafter\def\csname LT1\endcsname{\color[rgb]{0,1,0}}%
      \expandafter\def\csname LT2\endcsname{\color[rgb]{0,0,1}}%
      \expandafter\def\csname LT3\endcsname{\color[rgb]{1,0,1}}%
      \expandafter\def\csname LT4\endcsname{\color[rgb]{0,1,1}}%
      \expandafter\def\csname LT5\endcsname{\color[rgb]{1,1,0}}%
      \expandafter\def\csname LT6\endcsname{\color[rgb]{0,0,0}}%
      \expandafter\def\csname LT7\endcsname{\color[rgb]{1,0.3,0}}%
      \expandafter\def\csname LT8\endcsname{\color[rgb]{0.5,0.5,0.5}}%
    \else
      % gray
      \def\colorrgb#1{\color{black}}%
      \def\colorgray#1{\color[gray]{#1}}%
      \expandafter\def\csname LTw\endcsname{\color{white}}%
      \expandafter\def\csname LTb\endcsname{\color{black}}%
      \expandafter\def\csname LTa\endcsname{\color{black}}%
      \expandafter\def\csname LT0\endcsname{\color{black}}%
      \expandafter\def\csname LT1\endcsname{\color{black}}%
      \expandafter\def\csname LT2\endcsname{\color{black}}%
      \expandafter\def\csname LT3\endcsname{\color{black}}%
      \expandafter\def\csname LT4\endcsname{\color{black}}%
      \expandafter\def\csname LT5\endcsname{\color{black}}%
      \expandafter\def\csname LT6\endcsname{\color{black}}%
      \expandafter\def\csname LT7\endcsname{\color{black}}%
      \expandafter\def\csname LT8\endcsname{\color{black}}%
    \fi
  \fi
  \setlength{\unitlength}{0.0500bp}%
  \begin{picture}(5040.00,3528.00)%
    \gplgaddtomacro\gplbacktext{%
      \csname LTb\endcsname%
      \put(990,860){\makebox(0,0)[r]{\strut{} 1.2}}%
      \put(990,1261){\makebox(0,0)[r]{\strut{} 1.4}}%
      \put(990,1662){\makebox(0,0)[r]{\strut{} 1.6}}%
      \put(990,2062){\makebox(0,0)[r]{\strut{} 1.8}}%
      \put(990,2463){\makebox(0,0)[r]{\strut{} 2}}%
      \put(990,2863){\makebox(0,0)[r]{\strut{} 2.2}}%
      \put(990,3264){\makebox(0,0)[r]{\strut{} 2.4}}%
      \put(1122,440){\makebox(0,0){\strut{}-1}}%
      \put(1476,440){\makebox(0,0){\strut{}-0.8}}%
      \put(1831,440){\makebox(0,0){\strut{}-0.6}}%
      \put(2185,440){\makebox(0,0){\strut{}-0.4}}%
      \put(2540,440){\makebox(0,0){\strut{}-0.2}}%
      \put(2894,440){\makebox(0,0){\strut{} 0}}%
      \put(3248,440){\makebox(0,0){\strut{} 0.2}}%
      \put(3603,440){\makebox(0,0){\strut{} 0.4}}%
      \put(3957,440){\makebox(0,0){\strut{} 0.6}}%
      \put(4312,440){\makebox(0,0){\strut{} 0.8}}%
      \put(4666,440){\makebox(0,0){\strut{} 1}}%
      \put(220,1962){\rotatebox{90}{\makebox(0,0){\strut{}Exponent}}}%
      \put(2894,110){\makebox(0,0){\strut{}Shift}}%
    }%
    \gplgaddtomacro\gplfronttext{%
      \csname LTb\endcsname%
      \put(2574,3091){\makebox(0,0)[r]{\strut{}$\hat\alpha_{\mathrm{CSN}}$}}%
      \csname LTb\endcsname%
      \put(2574,2871){\makebox(0,0)[r]{\strut{}$\ev{\hat\alpha_{\mathrm{CSN}}}$}}%
      \csname LTb\endcsname%
      \put(2574,2651){\makebox(0,0)[r]{\strut{}$\hat\alpha$}}%
      \csname LTb\endcsname%
      \put(2574,2431){\makebox(0,0)[r]{\strut{}$\ev{\hat\alpha}$}}%
      \csname LTb\endcsname%
      \put(2574,2211){\makebox(0,0)[r]{\strut{}true value (1.5)}}%
    }%
    \gplbacktext
    \put(0,0){\includegraphics{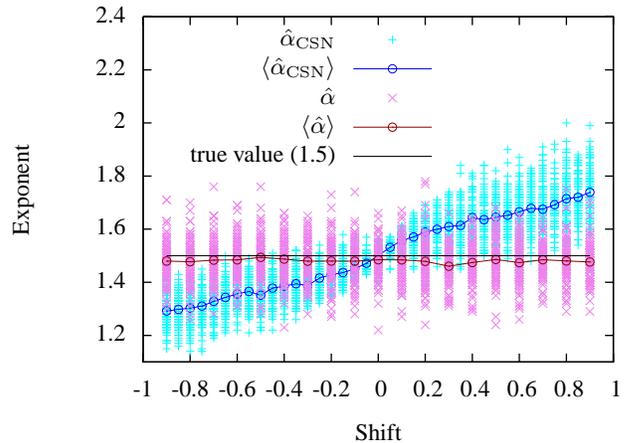}}%
    \gplfronttext
  \end{picture}%
\endgroup
 	\caption{Comparison of the CSNE estimator and the method introduced in Section \ref{sec:OptimizingTheShift}. Note that the standard deviation obtained using this new procedure is larger than that of the CSNE at zero shift, this is the price of shift-independence. (color online)}
	\label{fig:comparison}
 \end{figure}

\begin{figure}[tb]
	\centering
		%\input{1.50-test_shfit-1.tl.tex}
% GNUPLOT: LaTeX picture with Postscript
\begingroup
  \makeatletter
  \providecommand\color[2][]{%
    \GenericError{(gnuplot) \space\space\space\@spaces}{%
      Package color not loaded in conjunction with
      terminal option `colourtext'%
    }{See the gnuplot documentation for explanation.%
    }{Either use 'blacktext' in gnuplot or load the package
      color.sty in LaTeX.}%
    \renewcommand\color[2][]{}%
  }%
  \providecommand\includegraphics[2][]{%
    \GenericError{(gnuplot) \space\space\space\@spaces}{%
      Package graphicx or graphics not loaded%
    }{See the gnuplot documentation for explanation.%
    }{The gnuplot epslatex terminal needs graphicx.sty or graphics.sty.}%
    \renewcommand\includegraphics[2][]{}%
  }%
  \providecommand\rotatebox[2]{#2}%
  \@ifundefined{ifGPcolor}{%
    \newif\ifGPcolor
    \GPcolortrue
  }{}%
  \@ifundefined{ifGPblacktext}{%
    \newif\ifGPblacktext
    \GPblacktextfalse
  }{}%
  % define a \g@addto@macro without @ in the name:
  \let\gplgaddtomacro\g@addto@macro
  % define empty templates for all commands taking text:
  \gdef\gplbacktext{}%
  \gdef\gplfronttext{}%
  \makeatother
  \ifGPblacktext
    % no textcolor at all
    \def\colorrgb#1{}%
    \def\colorgray#1{}%
  \else
    % gray or color?
    \ifGPcolor
      \def\colorrgb#1{\color[rgb]{#1}}%
      \def\colorgray#1{\color[gray]{#1}}%
      \expandafter\def\csname LTw\endcsname{\color{white}}%
      \expandafter\def\csname LTb\endcsname{\color{black}}%
      \expandafter\def\csname LTa\endcsname{\color{black}}%
      \expandafter\def\csname LT0\endcsname{\color[rgb]{1,0,0}}%
      \expandafter\def\csname LT1\endcsname{\color[rgb]{0,1,0}}%
      \expandafter\def\csname LT2\endcsname{\color[rgb]{0,0,1}}%
      \expandafter\def\csname LT3\endcsname{\color[rgb]{1,0,1}}%
      \expandafter\def\csname LT4\endcsname{\color[rgb]{0,1,1}}%
      \expandafter\def\csname LT5\endcsname{\color[rgb]{1,1,0}}%
      \expandafter\def\csname LT6\endcsname{\color[rgb]{0,0,0}}%
      \expandafter\def\csname LT7\endcsname{\color[rgb]{1,0.3,0}}%
      \expandafter\def\csname LT8\endcsname{\color[rgb]{0.5,0.5,0.5}}%
    \else
      % gray
      \def\colorrgb#1{\color{black}}%
      \def\colorgray#1{\color[gray]{#1}}%
      \expandafter\def\csname LTw\endcsname{\color{white}}%
      \expandafter\def\csname LTb\endcsname{\color{black}}%
      \expandafter\def\csname LTa\endcsname{\color{black}}%
      \expandafter\def\csname LT0\endcsname{\color{black}}%
      \expandafter\def\csname LT1\endcsname{\color{black}}%
      \expandafter\def\csname LT2\endcsname{\color{black}}%
      \expandafter\def\csname LT3\endcsname{\color{black}}%
      \expandafter\def\csname LT4\endcsname{\color{black}}%
      \expandafter\def\csname LT5\endcsname{\color{black}}%
      \expandafter\def\csname LT6\endcsname{\color{black}}%
      \expandafter\def\csname LT7\endcsname{\color{black}}%
      \expandafter\def\csname LT8\endcsname{\color{black}}%
    \fi
  \fi
  \setlength{\unitlength}{0.0500bp}%
  \begin{picture}(5040.00,3528.00)%
    \gplgaddtomacro\gplbacktext{%
      \csname LTb\endcsname%
      \put(858,1731){\makebox(0,0)[r]{\strut{}$10^3$}}%
      \put(858,3264){\makebox(0,0)[r]{\strut{}$10^4$}}%
      \put(990,440){\makebox(0,0){\strut{}-1}}%
      \put(1358,440){\makebox(0,0){\strut{}-0.8}}%
      \put(1725,440){\makebox(0,0){\strut{}-0.6}}%
      \put(2093,440){\makebox(0,0){\strut{}-0.4}}%
      \put(2460,440){\makebox(0,0){\strut{}-0.2}}%
      \put(2828,440){\makebox(0,0){\strut{} 0}}%
      \put(3196,440){\makebox(0,0){\strut{} 0.2}}%
      \put(3563,440){\makebox(0,0){\strut{} 0.4}}%
      \put(3931,440){\makebox(0,0){\strut{} 0.6}}%
      \put(4298,440){\makebox(0,0){\strut{} 0.8}}%
      \put(4666,440){\makebox(0,0){\strut{} 1}}%
      \put(220,1962){\rotatebox{90}{\makebox(0,0){\strut{}Tail length}}}%
      \put(2828,110){\makebox(0,0){\strut{}Shift}}%
    }%
    \gplgaddtomacro\gplfronttext{%
      \csname LTb\endcsname%
      \put(3679,1493){\makebox(0,0)[r]{\strut{}$\hat n_{\mathrm{CSN}}$}}%
      \csname LTb\endcsname%
      \put(3679,1273){\makebox(0,0)[r]{\strut{}$\ev{\hat n_{\mathrm{CSN}}}$}}%
      \csname LTb\endcsname%
      \put(3679,1053){\makebox(0,0)[r]{\strut{}$\hat n$}}%
      \csname LTb\endcsname%
      \put(3679,833){\makebox(0,0)[r]{\strut{}$\ev{\hat n}$}}%
    }%
    \gplbacktext
    \put(0,0){\includegraphics{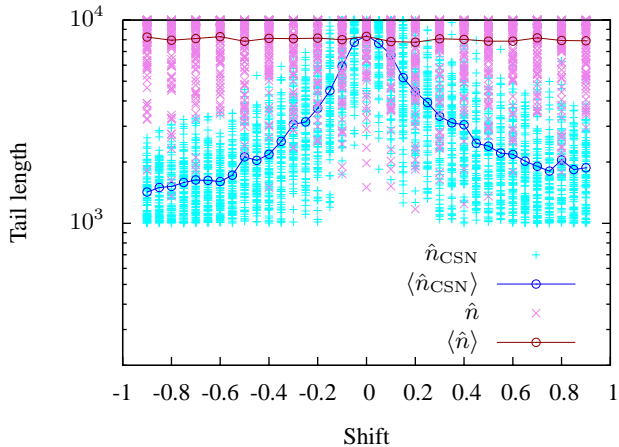}}%
    \gplfronttext
  \end{picture}%
\endgroup

	\caption{The estimated tail sizes in the procedure introduced in Section \ref{sec:OptimizingTheShift}. (color online)}
	\label{fig:ntail1}
\end{figure}

The procedure was tested on computer-generated datasets of size $N=10^4$ consisting of independent elements distributed according to a shifted power-law:
\begin{equation}
	\prob{X_i \geqslant x} = \left\{
\begin{array}{cl}
\br{x+s}^{-\alpha} & \text{if } x \geqslant 1,\\
1 & \text{otherwise,}\\
\end{array} \right.
	\label{eq:testdistr}
\end{equation}
for $i = 1, 2, \dots, N$. The parameters $s$ and $\alpha$ were varied in the range
\begin{eqnarray*}
	s & \in & [-0.9,0.9],\\
	\alpha & \in & \BR{1.5,2,2.5}. 
\end{eqnarray*}
Figures \ref{fig:comparison}--\ref{fig:ntail1} show the output of the procedure introduced in Section \ref{sec:OptimizingTheShift} for datasets with a fixed exponent $\alpha = 1.5$. One can conclude that the method accounts for the shift problem, although at the price of an increased standard deviation relative to the CSNE zero shift case. The estimates of the shift and the tail length are not as accurate as those of the exponent, but this is not surprising, as the tail-exponent is relatively stable to small changes in the shift and the tail length as well.

Table \ref{tab:mean_stdev} shows the performance of the new method for different $\alpha$ values, the averages comprise the estimates with all shift-values considered. The accuracy gets worse with increasing exponent, this is no surprise, since the shift estimator and the Hill estimator both display this property.

\begin{table}
 \centering
 \begin{tabular}{c|c|c}
  $\alpha$ & $\ev{\hat\alpha}$ & $\sqrt{\ev{\Delta\hat\alpha^2}}$ \\
  \hline
  \hline
  1.5 & 1.48 & 0.08 \\
  2.0 & 1.97 & 0.11 \\
  2.5 & 2.46 & 0.16
 \end{tabular}
 \caption{Average and standard deviation of the exponents estimated with the new method.}
 \label{tab:mean_stdev}
\end{table}

\section{Empirical data}
	\label{sec:EmpiricalData}
\begin{figure}[tb]
	\centering
		%\input{hist.tex}
% GNUPLOT: LaTeX picture with Postscript
\begingroup
  \makeatletter
  \providecommand\color[2][]{%
    \GenericError{(gnuplot) \space\space\space\@spaces}{%
      Package color not loaded in conjunction with
      terminal option `colourtext'%
    }{See the gnuplot documentation for explanation.%
    }{Either use 'blacktext' in gnuplot or load the package
      color.sty in LaTeX.}%
    \renewcommand\color[2][]{}%
  }%
  \providecommand\includegraphics[2][]{%
    \GenericError{(gnuplot) \space\space\space\@spaces}{%
      Package graphicx or graphics not loaded%
    }{See the gnuplot documentation for explanation.%
    }{The gnuplot epslatex terminal needs graphicx.sty or graphics.sty.}%
    \renewcommand\includegraphics[2][]{}%
  }%
  \providecommand\rotatebox[2]{#2}%
  \@ifundefined{ifGPcolor}{%
    \newif\ifGPcolor
    \GPcolortrue
  }{}%
  \@ifundefined{ifGPblacktext}{%
    \newif\ifGPblacktext
    \GPblacktexttrue
  }{}%
  % define a \g@addto@macro without @ in the name:
  \let\gplgaddtomacro\g@addto@macro
  % define empty templates for all commands taking text:
  \gdef\gplbacktext{}%
  \gdef\gplfronttext{}%
  \makeatother
  \ifGPblacktext
    % no textcolor at all
    \def\colorrgb#1{}%
    \def\colorgray#1{}%
  \else
    % gray or color?
    \ifGPcolor
      \def\colorrgb#1{\color[rgb]{#1}}%
      \def\colorgray#1{\color[gray]{#1}}%
      \expandafter\def\csname LTw\endcsname{\color{white}}%
      \expandafter\def\csname LTb\endcsname{\color{black}}%
      \expandafter\def\csname LTa\endcsname{\color{black}}%
      \expandafter\def\csname LT0\endcsname{\color[rgb]{1,0,0}}%
      \expandafter\def\csname LT1\endcsname{\color[rgb]{0,1,0}}%
      \expandafter\def\csname LT2\endcsname{\color[rgb]{0,0,1}}%
      \expandafter\def\csname LT3\endcsname{\color[rgb]{1,0,1}}%
      \expandafter\def\csname LT4\endcsname{\color[rgb]{0,1,1}}%
      \expandafter\def\csname LT5\endcsname{\color[rgb]{1,1,0}}%
      \expandafter\def\csname LT6\endcsname{\color[rgb]{0,0,0}}%
      \expandafter\def\csname LT7\endcsname{\color[rgb]{1,0.3,0}}%
      \expandafter\def\csname LT8\endcsname{\color[rgb]{0.5,0.5,0.5}}%
    \else
      % gray
      \def\colorrgb#1{\color{black}}%
      \def\colorgray#1{\color[gray]{#1}}%
      \expandafter\def\csname LTw\endcsname{\color{white}}%
      \expandafter\def\csname LTb\endcsname{\color{black}}%
      \expandafter\def\csname LTa\endcsname{\color{black}}%
      \expandafter\def\csname LT0\endcsname{\color{black}}%
      \expandafter\def\csname LT1\endcsname{\color{black}}%
      \expandafter\def\csname LT2\endcsname{\color{black}}%
      \expandafter\def\csname LT3\endcsname{\color{black}}%
      \expandafter\def\csname LT4\endcsname{\color{black}}%
      \expandafter\def\csname LT5\endcsname{\color{black}}%
      \expandafter\def\csname LT6\endcsname{\color{black}}%
      \expandafter\def\csname LT7\endcsname{\color{black}}%
      \expandafter\def\csname LT8\endcsname{\color{black}}%
    \fi
  \fi
  \setlength{\unitlength}{0.0500bp}%
  \begin{picture}(4896.00,3426.00)%
    \gplgaddtomacro\gplbacktext{%
      \csname LTb\endcsname%
      \put(990,660){\makebox(0,0)[r]{\strut{} 0}}%
      \put(990,1160){\makebox(0,0)[r]{\strut{} 50}}%
      \put(990,1661){\makebox(0,0)[r]{\strut{} 100}}%
      \put(990,2161){\makebox(0,0)[r]{\strut{} 150}}%
      \put(990,2662){\makebox(0,0)[r]{\strut{} 200}}%
      \put(990,3162){\makebox(0,0)[r]{\strut{} 250}}%
      \put(1122,440){\makebox(0,0){\strut{} 0}}%
      \put(1547,440){\makebox(0,0){\strut{} 0.5}}%
      \put(1972,440){\makebox(0,0){\strut{} 1}}%
      \put(2397,440){\makebox(0,0){\strut{} 1.5}}%
      \put(2822,440){\makebox(0,0){\strut{} 2}}%
      \put(3247,440){\makebox(0,0){\strut{} 2.5}}%
      \put(3672,440){\makebox(0,0){\strut{} 3}}%
      \put(4097,440){\makebox(0,0){\strut{} 3.5}}%
      \put(4522,440){\makebox(0,0){\strut{} 4}}%
      \put(220,1911){\rotatebox{90}{\makebox(0,0){\strut{}Number of occurrences}}}%
      \put(2822,110){\makebox(0,0){\strut{}Exponent}}%
    }%
    \gplgaddtomacro\gplfronttext{%
      \csname LTb\endcsname%
      \put(3535,2989){\makebox(0,0)[r]{\strut{}CSNE fit}}%
      \csname LTb\endcsname%
      \put(3535,2769){\makebox(0,0)[r]{\strut{}Sh.-opt. fit}}%
    }%
    \gplbacktext
    \put(0,0){\includegraphics{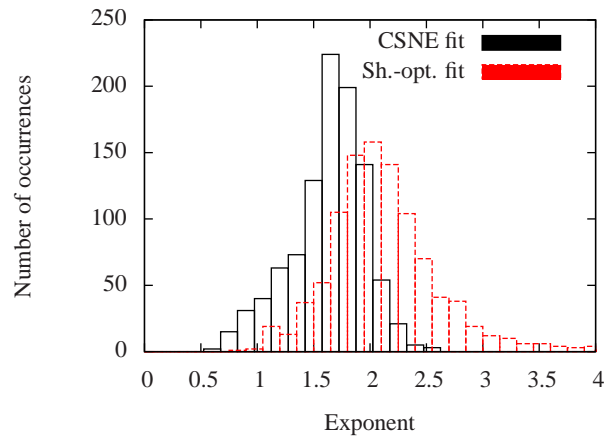}}%
    \gplfronttext
  \end{picture}%
\endgroup

	\caption{Comparison of the CSNE and the new procedure on empirical data. The black bars correspond to the histogram of the CSNE estimates, the red dashed bars to the new one. (color online)}
	\label{fig:shfreal}
\end{figure}

\begin{figure}[tb]
	\centering
	\begin{tabular}{c}
		%\input{KM.linlog.tex}
% GNUPLOT: LaTeX picture with Postscript
\begingroup
  \makeatletter
  \providecommand\color[2][]{%
    \GenericError{(gnuplot) \space\space\space\@spaces}{%
      Package color not loaded in conjunction with
      terminal option `colourtext'%
    }{See the gnuplot documentation for explanation.%
    }{Either use 'blacktext' in gnuplot or load the package
      color.sty in LaTeX.}%
    \renewcommand\color[2][]{}%
  }%
  \providecommand\includegraphics[2][]{%
    \GenericError{(gnuplot) \space\space\space\@spaces}{%
      Package graphicx or graphics not loaded%
    }{See the gnuplot documentation for explanation.%
    }{The gnuplot epslatex terminal needs graphicx.sty or graphics.sty.}%
    \renewcommand\includegraphics[2][]{}%
  }%
  \providecommand\rotatebox[2]{#2}%
  \@ifundefined{ifGPcolor}{%
    \newif\ifGPcolor
    \GPcolortrue
  }{}%
  \@ifundefined{ifGPblacktext}{%
    \newif\ifGPblacktext
    \GPblacktextfalse
  }{}%
  % define a \g@addto@macro without @ in the name:
  \let\gplgaddtomacro\g@addto@macro
  % define empty templates for all commands taking text:
  \gdef\gplbacktext{}%
  \gdef\gplfronttext{}%
  \makeatother
  \ifGPblacktext
    % no textcolor at all
    \def\colorrgb#1{}%
    \def\colorgray#1{}%
  \else
    % gray or color?
    \ifGPcolor
      \def\colorrgb#1{\color[rgb]{#1}}%
      \def\colorgray#1{\color[gray]{#1}}%
      \expandafter\def\csname LTw\endcsname{\color{white}}%
      \expandafter\def\csname LTb\endcsname{\color{black}}%
      \expandafter\def\csname LTa\endcsname{\color{black}}%
      \expandafter\def\csname LT0\endcsname{\color[rgb]{1,0,0}}%
      \expandafter\def\csname LT1\endcsname{\color[rgb]{0,1,0}}%
      \expandafter\def\csname LT2\endcsname{\color[rgb]{0,0,1}}%
      \expandafter\def\csname LT3\endcsname{\color[rgb]{1,0,1}}%
      \expandafter\def\csname LT4\endcsname{\color[rgb]{0,1,1}}%
      \expandafter\def\csname LT5\endcsname{\color[rgb]{1,1,0}}%
      \expandafter\def\csname LT6\endcsname{\color[rgb]{0,0,0}}%
      \expandafter\def\csname LT7\endcsname{\color[rgb]{1,0.3,0}}%
      \expandafter\def\csname LT8\endcsname{\color[rgb]{0.5,0.5,0.5}}%
    \else
      % gray
      \def\colorrgb#1{\color{black}}%
      \def\colorgray#1{\color[gray]{#1}}%
      \expandafter\def\csname LTw\endcsname{\color{white}}%
      \expandafter\def\csname LTb\endcsname{\color{black}}%
      \expandafter\def\csname LTa\endcsname{\color{black}}%
      \expandafter\def\csname LT0\endcsname{\color{black}}%
      \expandafter\def\csname LT1\endcsname{\color{black}}%
      \expandafter\def\csname LT2\endcsname{\color{black}}%
      \expandafter\def\csname LT3\endcsname{\color{black}}%
      \expandafter\def\csname LT4\endcsname{\color{black}}%
      \expandafter\def\csname LT5\endcsname{\color{black}}%
      \expandafter\def\csname LT6\endcsname{\color{black}}%
      \expandafter\def\csname LT7\endcsname{\color{black}}%
      \expandafter\def\csname LT8\endcsname{\color{black}}%
    \fi
  \fi
  \setlength{\unitlength}{0.0500bp}%
  \begin{picture}(4750.00,3326.00)%
    \gplgaddtomacro\gplbacktext{%
      \csname LTb\endcsname%
      \put(990,878){\makebox(0,0)[r]{\strut{} 0}}%
      \put(990,1315){\makebox(0,0)[r]{\strut{} 0.2}}%
      \put(990,1752){\makebox(0,0)[r]{\strut{} 0.4}}%
      \put(990,2189){\makebox(0,0)[r]{\strut{} 0.6}}%
      \put(990,2625){\makebox(0,0)[r]{\strut{} 0.8}}%
      \put(990,3062){\makebox(0,0)[r]{\strut{} 1}}%
      \put(1122,440){\makebox(0,0){\strut{} 1}}%
      \put(1664,440){\makebox(0,0){\strut{}10}}%
      \put(2207,440){\makebox(0,0){\strut{} $10^2$}}%
      \put(2749,440){\makebox(0,0){\strut{} $10^3$}}%
      \put(3291,440){\makebox(0,0){\strut{} $10^4$}}%
      \put(3834,440){\makebox(0,0){\strut{} $10^5$}}%
      \put(4376,440){\makebox(0,0){\strut{} $10^6$}}%
      \put(220,1861){\rotatebox{90}{\makebox(0,0){\strut{}CDF}}}%
      \put(2749,110){\makebox(0,0){\strut{}Trade volume (USD)}}%
    }%
    \gplgaddtomacro\gplfronttext{%
      \csname LTb\endcsname%
      \put(3389,2889){\makebox(0,0)[r]{\strut{}empirical cdf, KM}}%
      \csname LTb\endcsname%
      \put(3389,2669){\makebox(0,0)[r]{\strut{}CSNE fit}}%
      \csname LTb\endcsname%
      \put(3389,2449){\makebox(0,0)[r]{\strut{}Shift-optimized fit}}%
    }%
    \gplbacktext
    \put(0,0){\includegraphics{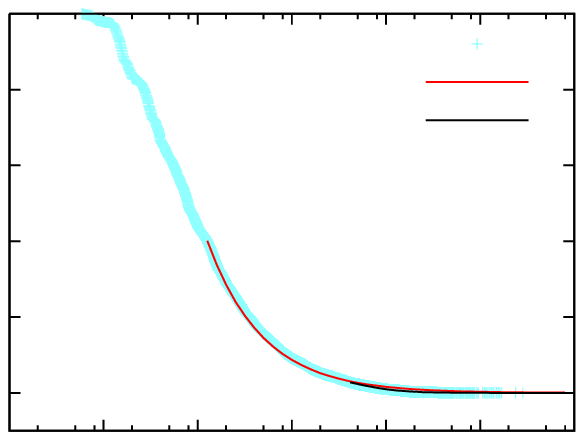}}%
    \gplfronttext
  \end{picture}%
\endgroup

		\\

%\input{KM.loglog.tex}
% GNUPLOT: LaTeX picture with Postscript
\begingroup
  \makeatletter
  \providecommand\color[2][]{%
    \GenericError{(gnuplot) \space\space\space\@spaces}{%
      Package color not loaded in conjunction with
      terminal option `colourtext'%
    }{See the gnuplot documentation for explanation.%
    }{Either use 'blacktext' in gnuplot or load the package
      color.sty in LaTeX.}%
    \renewcommand\color[2][]{}%
  }%
  \providecommand\includegraphics[2][]{%
    \GenericError{(gnuplot) \space\space\space\@spaces}{%
      Package graphicx or graphics not loaded%
    }{See the gnuplot documentation for explanation.%
    }{The gnuplot epslatex terminal needs graphicx.sty or graphics.sty.}%
    \renewcommand\includegraphics[2][]{}%
  }%
  \providecommand\rotatebox[2]{#2}%
  \@ifundefined{ifGPcolor}{%
    \newif\ifGPcolor
    \GPcolortrue
  }{}%
  \@ifundefined{ifGPblacktext}{%
    \newif\ifGPblacktext
    \GPblacktextfalse
  }{}%
  % define a \g@addto@macro without @ in the name:
  \let\gplgaddtomacro\g@addto@macro
  % define empty templates for all commands taking text:
  \gdef\gplbacktext{}%
  \gdef\gplfronttext{}%
  \makeatother
  \ifGPblacktext
    % no textcolor at all
    \def\colorrgb#1{}%
    \def\colorgray#1{}%
  \else
    % gray or color?
    \ifGPcolor
      \def\colorrgb#1{\color[rgb]{#1}}%
      \def\colorgray#1{\color[gray]{#1}}%
      \expandafter\def\csname LTw\endcsname{\color{white}}%
      \expandafter\def\csname LTb\endcsname{\color{black}}%
      \expandafter\def\csname LTa\endcsname{\color{black}}%
      \expandafter\def\csname LT0\endcsname{\color[rgb]{1,0,0}}%
      \expandafter\def\csname LT1\endcsname{\color[rgb]{0,1,0}}%
      \expandafter\def\csname LT2\endcsname{\color[rgb]{0,0,1}}%
      \expandafter\def\csname LT3\endcsname{\color[rgb]{1,0,1}}%
      \expandafter\def\csname LT4\endcsname{\color[rgb]{0,1,1}}%
      \expandafter\def\csname LT5\endcsname{\color[rgb]{1,1,0}}%
      \expandafter\def\csname LT6\endcsname{\color[rgb]{0,0,0}}%
      \expandafter\def\csname LT7\endcsname{\color[rgb]{1,0.3,0}}%
      \expandafter\def\csname LT8\endcsname{\color[rgb]{0.5,0.5,0.5}}%
    \else
      % gray
      \def\colorrgb#1{\color{black}}%
      \def\colorgray#1{\color[gray]{#1}}%
      \expandafter\def\csname LTw\endcsname{\color{white}}%
      \expandafter\def\csname LTb\endcsname{\color{black}}%
      \expandafter\def\csname LTa\endcsname{\color{black}}%
      \expandafter\def\csname LT0\endcsname{\color{black}}%
      \expandafter\def\csname LT1\endcsname{\color{black}}%
      \expandafter\def\csname LT2\endcsname{\color{black}}%
      \expandafter\def\csname LT3\endcsname{\color{black}}%
      \expandafter\def\csname LT4\endcsname{\color{black}}%
      \expandafter\def\csname LT5\endcsname{\color{black}}%
      \expandafter\def\csname LT6\endcsname{\color{black}}%
      \expandafter\def\csname LT7\endcsname{\color{black}}%
      \expandafter\def\csname LT8\endcsname{\color{black}}%
    \fi
  \fi
  \setlength{\unitlength}{0.0500bp}%
  \begin{picture}(4750.00,3326.00)%
    \gplgaddtomacro\gplbacktext{%
      \csname LTb\endcsname%
      \put(990,660){\makebox(0,0)[r]{\strut{} $10^{-7}$}}%
      \put(990,1003){\makebox(0,0)[r]{\strut{} $10^{-6}$}}%
      \put(990,1346){\makebox(0,0)[r]{\strut{} $10^{-5}$}}%
      \put(990,1689){\makebox(0,0)[r]{\strut{} $10^{-4}$}}%
      \put(990,2033){\makebox(0,0)[r]{\strut{} $10^{-3}$}}%
      \put(990,2376){\makebox(0,0)[r]{\strut{} $10^{-2}$}}%
      \put(990,2719){\makebox(0,0)[r]{\strut{} $10^{-1}$}}%
      \put(990,3062){\makebox(0,0)[r]{\strut{} 1}}%
      \put(1122,440){\makebox(0,0){\strut{} 1}}%
      \put(1664,440){\makebox(0,0){\strut{} 10}}%
      \put(2207,440){\makebox(0,0){\strut{} $10^2$}}%
      \put(2749,440){\makebox(0,0){\strut{} $10^3$}}%
      \put(3291,440){\makebox(0,0){\strut{} $10^4$}}%
      \put(3834,440){\makebox(0,0){\strut{} $10^5$}}%
      \put(4376,440){\makebox(0,0){\strut{} $10^6$}}%
      \put(220,1861){\rotatebox{90}{\makebox(0,0){\strut{}CDF}}}%
      \put(2749,110){\makebox(0,0){\strut{}Trade volume (USD)}}%
    }%
    \gplgaddtomacro\gplfronttext{%
      \csname LTb\endcsname%
      \put(3498,1273){\makebox(0,0)[r]{\strut{}empirical cdf, KM}}%
      \csname LTb\endcsname%
      \put(3498,1053){\makebox(0,0)[r]{\strut{}CSNE fit}}%
      \csname LTb\endcsname%
      \put(3498,833){\makebox(0,0)[r]{\strut{}Shift-optimized fit}}%
    }%
    \gplbacktext
    \put(0,0){\includegraphics{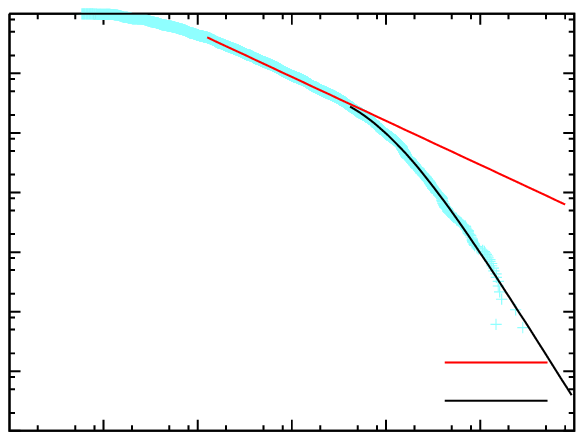}}%
    \gplfronttext
  \end{picture}%
\endgroup

	\end{tabular}
	\caption{The cdf of trade sizes for the shares of the Kmart Corporation (cyan crosses). The red and black curves show the fits provided by the two estimation procedures. The upper and lower plots differ only in the scale of the vertical axis. (color online)}
	\label{fig:example}
\end{figure}

As an application, we use the method to determine tail exponents of stock market data.
 The dataset considered was that of the dollar value \footnote{We measured trade volumes in US dollars to avoid the problem of stock splits.} of individual transactions of the 1000 largest stocks (according to the total number of trades in the period) in 1994-95 at the New York Stock Exchange \cite{taq}.
 We wish to emphasize, however, as there is no firm theoretical evidence supporting the power-law property of the distribution of trade sizes, that we see it only as one possibility to model the tail of the cdf.
 Fig.\ \ref{fig:shfreal} compares the histogram of the tail exponent estimates gained using the CSNE estimator, and the new, shift-invariant procedure.
 Fig.\ \ref{fig:example} shows the fits provided by the two estimators on a stock for which they provide very different estimates.

The new procedure found fits with a KS distance on average 23\% lower than the CSNE estimator. This is not a drastic difference, but on a logarithmic scale, the KS statistic is less sensitive in the tail than for small $x$ values. Fig.\ \ref{fig:example} demonstrates this in the case of a specific stock (Kmart Corporation): although on a linear scale (upper part), the two fits do not seem to be very different, the logarithmic plot (lower part) shows that the two deviate in the tail region. In this specific case, the distance of the new fit improved from 0.035 to 0.021. %This observation leads us to believe that there might be a better notion of distance to use than the KS statistic, however, modifying the KS to work with logarithmic differences does not produce much different exponents. 

In empirical data, an additional problem has to be considered when analyzing the results. In case there is no strong theoretical indication for Eq.\ \eqref{eq:heavytails2}, this type of ansatz can only be accepted, if there are at least some datapoints in the tail, i.\ e., for which the approximation
\begin{equation}
 \frac{x^{-\alpha}}{\br{x+s}^{-\alpha}} \approx 1 + \frac{\alpha \cdot s}{x}
 \label{eq:approx}
\end{equation}
is applicable. This is of importance because fitting an exponent to a non-observed tail is questionable, and even if there is theoretical evidence for a shifted power-law tail, errors are amplified. In other words, the quantity
\begin{equation}
 \delta = \frac{\hat \alpha \cdot \hat s}{x_{\max}},
 \label{eq:delta}
\end{equation}
(with $x_{\max} = \max_i \left\{X_i \in \mathcal S\right\}$) which measures the ``distance'' of the largest observation from the tail region, has to be small. For the data analyzed, about 10\% of the stocks had $\delta \geqslant 0.1$. If we exclude these data from the statistic, the average exponent is 2.0 with a standard deviation of 0.35 \footnote{Fig.\ \ref{fig:shfreal} is the histogram of all obtained exponents, $\delta \geqslant 0.1$ included. The average is in the non-filtered case higher, 2.13.}.

One can conclude that in the case of stock trade volumes, the inclusion of data shifts clearly has an effect on the results. Furthermore, note that the typical value of the estimates is approximately 2, i.\ e.\, on the boundary of the Levy regime. 
This matter has been controversial and our present results support the view that the exponents are higher than thought earlier (Refs.\ \cite{Gopikrishnan00},\cite{Gabaix03a},\cite{Eisler06} and \cite{Plerou07}). Furthermore, since the estimates have a large standard deviation, we find that the term universality is not applicable to trade volume distributions. 

\section{Conclusions}
	\label{sec:Conclusions}
In this paper, we have shown for empirical as well as computer-generated data, that data shifts can play an important role in the tail exponent estimation procedure.
 Tests on computer-generated datasets showed that this problem can be solved by the suggested extension of the Hill estimator. A small, yet systematic underestimation of the exponent is present, nevertheless, on empirical data, this bias loses its importance when compared to other error sources.
 Our results regarding stock market data lead us to doubts about the idea of universal tail exponents (Ref.\ \cite{Gopikrishnan00}-\cite{Gabaix03a}) regarding both universality and typical values.

\appendix*
\section{The CSNE algorithm \cite{Clauset07}}
	\label{sec:Appendix}
%Clauset estimator => for the appendix
The CSNE method optimizes the tail length parameter of the Hill estimator: the algorithm calculates $\hat \alpha_{\mathrm H}\br{\mathcal S, n}$ for many different $n$-s and accepts the fit which is the closest to the empirical cdf, according to the Kolmogorov--Smirnov statistic.
In terms of formulas, the procedure for a given set $\mathcal S = \BR{X_1, X_2, \dots, X_N}$ observations, can be described as follows:
\begin{enumerate}
	\item Sorting:
		\begin{eqnarray*}
			& \BR{X_1, \dots, X_N} \rightarrow \BR{X_{\br 1}, \dots, X_{\br N}}: & \\
			& X_{\br 1} \geqslant X_{\br 2} \geqslant \dots \geqslant X_{\br N} & 
			\label{eq:sorting}
		\end{eqnarray*}
	\item Choose the set $\mathcal C \subseteq \BR{1, 2, \dots, N}$ of tail lenghts to check, according to the following criteria:
		\begin{itemize}
			\item Do not include $n\leqslant n_{\mathrm{min}}$, because it does not make any sense to calculate the Hill estimator based on e.\ g.\ $5$ elements.
			\item If $N - n_{\mathrm{min}} + 1 > M$ choose $M$ elements from $\{n_{\min},$ $ n_{\min}+1,$ $ \dots,$ $ N\} $ uniformly to obtain $\mathcal C$.
		\end{itemize}
	\item Distance: Kolmogorov--Smirnov statistic:
		\begin{eqnarray}
			\nonumber &&\tilde F_x \br{x_i} \equiv \mathbb P \br{X \geqslant x_i \mid X  \geqslant x} \\
			&&D_{\mathrm {KS}}\br x  = \mathop {\max }\limits_{x_i  \geqslant x}  \abs{\tilde F_x\br{x_i} - \br{\frac{x_i}
				{x}}^{ - \hat \alpha_{\mathrm H} } }
			\label{eq:KS}
		\end{eqnarray}
	\item For all $X_{\br n} \in \mathcal C$ calculate the Hill estimator $\hat \alpha_{\mathrm H}^{-1}\br{\mathcal S, n}$ and the distance of the Hill fit and the empirical distribution function, $D_{\mathrm{KS}}\br{n}$.
	\item Accept the tail-length $n$ which minimizes the distance:
		\begin{equation*}
			\hat n = \arg \mathop {\min }\limits_{n \in \mathcal C} D_{\mathrm{KS}}\br{X_{\br n}}
			\label{eq:xmin}
		\end{equation*}
	\item As a result, we obtain:
		\begin{itemize}
			\item $\hat \alpha = \hat \alpha_{\mathrm H} \br{\mathcal S, \hat n_{\mathrm{tail}}}$
			\item $\hat x_0 = X_{\br{\hat n}}$
			\item $d = D_{\mathrm{KS}}\br{\hat x_0}$
		\end{itemize}
\end{enumerate}

\bibliographystyle{unsrt}
\bibliography{shopt}

\end{document}